# One-sided destructive quantum interference from an exceptional point-enabled metasurface

Hong Liang[1], Kai Ming Lau[1], Wai Chun Wong[1], Shengwang Du[2], Wing Yim Tam[1], and Jensen Li[1*]

[1]Department of Physics, The Hong Kong University of Science and Technology, Clear Water Bay, Kowloon, Hong Kong, China.

[2]Department of Physics, The University of Texas at Dallas, Richardson, Texas 75080, USA

We propose the concept of one-sided quantum interference based on non-Hermitian metasurfaces. By designing bianisotropic metasurfaces with a non-Hermitian exceptional point, we show that quantum interference can exist only on only one side but not another. This is the quantum inheritance of unidirectional zero reflection in classical optics. The one-side interference can be further manipulated with tailor-made metasurface. With two photons simultaneously entering the metasurface from different sides, the probability for only outputting one photon on the side with reflection can be modified to zero as a one-sided destructive quantum interference while the output on another side is free of interference. We design the required bianisotropic metasurface and numerically demonstrate the proposed effect. The non-Hermitian bianisotropic metasurfaces provide more degrees of freedom in tuning two-photon quantum interference, in parallel to the celebrated Hong–Ou–Mandel effect.

---

[*] Corresponding author: jensenli@ust.hk

Metasurface, a thin layer of nanostructures array, is useful for miniaturizing conventional optical elements, with applications ranging from beam steering, lensing, to vortex beam structuring, holograms, and hybrid combinations of these functions [1-3]. Metasurfaces can also be used as a very flexible platform to study exotic physics, such as topological photonics and non-Hermitian physics, which explore robustness and configurational sensitivity in the optical properties [4,5]. For non-Hermitian metasurfaces, we exploit the role of material loss in the optical properties. The developments in the past decade in non-Hermitian photonics have found that the material loss, in competition with coupling between different parts of the system, can be used to generate an exceptional point: the coalesce of both the eigenvalues and eigenvectors of the system Hamiltonian or response matrix [6,7]. It then gives rise to a series of counter-intuitive wave phenomena, including loss-induced transmission [8], unidirectional zero reflection [9], enhanced sensing [10], lasing-mode selection [11], as some examples based on the configurational sensitivity of the exceptional points. These exceptional points have been demonstrated using metamaterials [12,13]. Specifically, unidirectional zero reflection (UZR) occurs at an exceptional point of the scattering matrix and can be used as a signature of a non-Hermitian exceptional point of metasurfaces [9,14,15]. With exceptional points at the optical frequency being recently observed [16,17,18], non-Hermitian physics is turning the undesired loss, which is often unavoidable in optical metasurfaces, to a useful tool in optical information processing and sensing.

In the quantum optical regime, tailor-made metasurfaces have already been found useful to simplify quantum tomography [19] and to generate entanglement for spins, orbital angular momenta, etc [20-22]. With the help of the enormous engineering degrees of freedom from the metasurfaces structures, metasurfaces pave a promising road to miniaturization of quantum optical elements. While conventional quantum optical operations [23-26] are constructed from a network of common optical components such as beam splitters, wave plates, metasurfaces allow us to design tailor-made unitary transformations for any specified unitary transformations. For example, arbitrary unitary U(2) operations and single-photon two-qubit U(4) operations can be realized using dielectric metasurfaces [27]. On the other hand, non-Hermitian metasurfaces (with material loss) also plays an important role to construct tailor-made non-unitary transformations, adding a new degree of freedom in tuning from destructive two-photon quantum interference to a constructive interference [28]. It is interesting to note that while the concept of non-Hermitian exceptional point has an origin in quantum theory [6,7], it was firstly realized in classical optical systems, and we start to see their realizations in quantum optical systems recently [29-31]. To further enable the non-Hermitian properties of metasurfaces, it is therefore natural to ask the role of exceptional point in manipulating two-photon quantum interference.

In this work, we propose a one-sided two-photon quantum interference effect for a two-port device by introducing non-Hermiticity, or material loss, into the system. By viewing a metasurface as a two-port device, similar to a beam-splitter in conventional quantum operations, we further propose a design of lossy bianisotropic metasurfaces capable for such quantum interference effect. By tailoring the resonances of the bianisotropic measurface, one-sided two-photon destructive quantum interference can be further achieved and numerically demonstrated through full-wave simulations with ancilla mode formulation and a coupled mode theory as guidance.

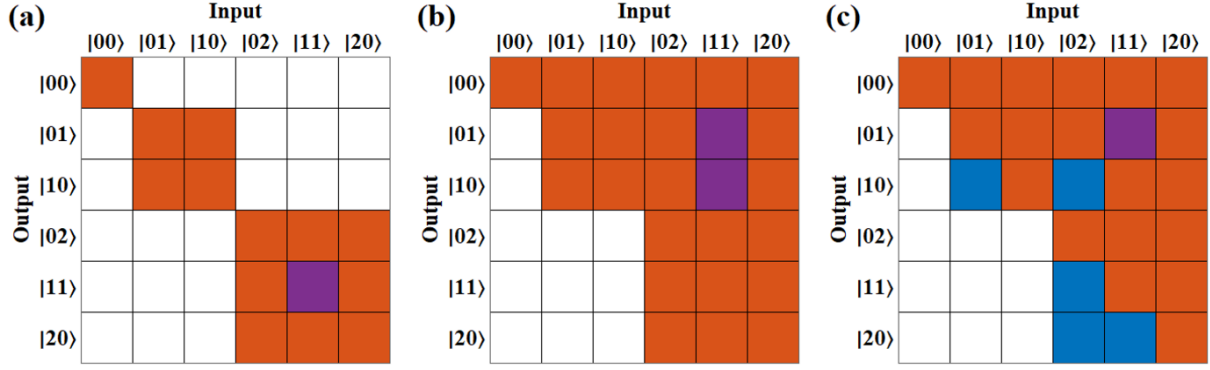

FIG. 1. Quantum interference for a two-port device. (a) HOM effect from a lossless 50/50 beam splitter: |11⟩ input state is forbidden to transfer to |11⟩ output through destructive quantum interference with zero probability, indicated as purple color. The white color denotes zero transfer probabilities due to conservation of energy while the orange color denotes generally non-zero values. (b). Apparent nonlinear absorption from a lossy system: |11⟩ input state is forbidden to transfer to single photon states |01⟩ and |10⟩ but not the states with 0 or 2 photons. (c) One-sided destructive quantum interference. |11⟩ input state is forbidden to transfer to |01⟩ state through quantum destructive interference but the quantum interference is off for output state |10⟩. It can be achieved based on UZR with zero backward reflection, indicated by the zero transfer probability from |01⟩ → |10⟩ and other processes with blue color.

Two-photon quantum interference effects for a two-port device can be visualized by an input-output map as shown in Fig. 2. Each square cell, labelled by its input state in column and output state in row, is colored orange for generally non-zero values, purple for zero from destructive quantum interference and white for zero from conservation of energy. Each input or output state is represented by a Fock state notation of two indices with first (second) index indicating the number of photons in port 1(2). For the celebrated Hong–Ou–Mandel (HOM) effect [32], the two ports are for a lossless 50/50 beam splitter and the process |11⟩ → |11⟩ is forbidden (purple color in Fig. 2(a)), which is a two-photon destructive quantum interference effect. When loss (non-Hermiticity) is added to a beam splitter, Fig. 2(b) shows the case of apparent nonlinear absorption enabled by a 25/25 beamsplitter [33-37]. For a two-photon input state |11⟩, the probabilities for only

one photon survives, $|10\rangle$ and $|01\rangle$, are both zero, i.e., either both photons are absorbed or neither is. Such a loss-induced quantum interference effect has been recently extended to plasmonic systems and metasurfaces, whose scattering matrix can now be tailor-made to achieve constructive quantum interference [28, 38]. The one-sided quantum interference effect proposed here is schematically illustrated in Fig. 2(c). For a $|11\rangle$ input, we would like the quantum interference effect exists only on one side to get a destructive interference at the $|01\rangle$ output while the quantum interference is turned off at the $|10\rangle$ output. In this case, we define the port 1 as forward and port 2 as backward normal incidence on the metasurfaces. As the metasurface is assumed reciprocal here, to get the interference effect on and off between $|01\rangle$ and $|10\rangle$, we would resort to a metasurface with the so-called unidirectional zero reflection (UZR), with the backward reflection coefficient being zero. Apart from the zero value at $|11\rangle \rightarrow |01\rangle$, we also expect zero values at $|01\rangle \rightarrow |10\rangle$ (UZR in single photon picture) and other processes where the number of photons in port 1 in output is larger than the one in input (blue color Fig. 2(c)).

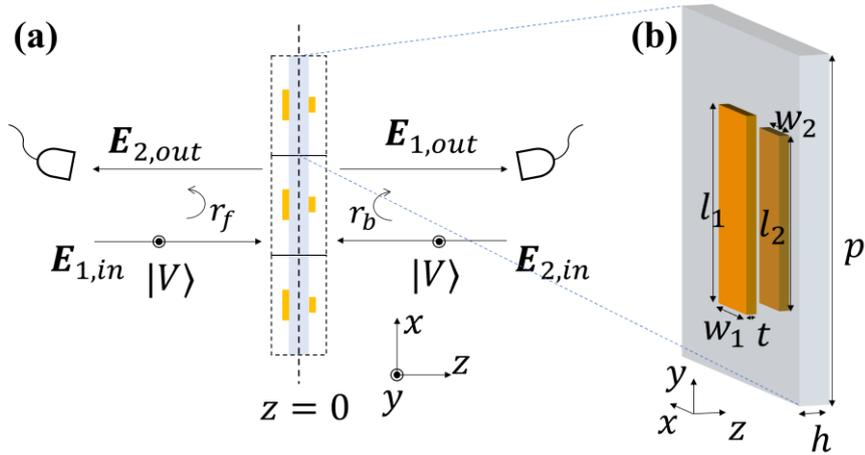

FIG. 2. (a) Proposed experiment with bianisotropic metasurface. $E_{1,in}$ ($E_{2,in}$) denotes the E-field, polarized along y-direction, at input port 1 (2). Subscript "out" is used for E-fields at output ports. (b) A square unit cell of metasurface with periodicity 300 nm, two vertical gold bars (thickness 20 nm) with dimensions $l_2$ = 150 nm, and $w_2$ = 50 nm while $l_1$, $w_1$ and separation $h$ will be swept in later designs. Drude model for the permittivity of gold: $\varepsilon = 9 - f_p^2/(f^2 + i\gamma_p f)$ with $f_p$ = 2180 THz, $\gamma_p$ = 16.2 THz.

Figure 2 shows our bianisotropic metasurface, located at $z = 0$, to have UZR as a starting point of consideration. Normal incidence from the front (back) side is defined as input port 1(2) while its transmitted light is defined as the corresponding output port. Each unit cell in a square lattice consists of two vertical gold bars with separation $h$, lengths $l_1$ and $l_2$, widths $w_1$ and $w_2$, and is driven into resonance by incident light polarized along the y-direction. Full-wave simulations (CST Studio Suite) are performed to obtain the classical response at each frequency, as a 2×2 scattering matrix $S = \{\{t_f, r_b\}, \{r_f, t_b\}\}$ where $t_f = t_b = t$ for reciprocity and $r_f \neq r_b$ for the broken mirror symmetry along z-direction. Subscript "f" ("b") stands for forward (backward) incidence. We scan $l_1$ (length of the front bar) and $h$ (separation between the two bars) with other dimensions listed in caption of Fig. 3 and $w_1$ is fixed at 80 nm. As shown in Fig. 3(a), we plot the minimum of $|r_b|$ around the resonating frequency at each coordinate $(l_1, h)$. The UZR condition at $|r_b| = 0$ is then shown up as the black curve in the phase space. We get a family of candidate metasurfaces by walking along the black curve. With a larger $l_1$, it is found that a smaller $h$ is needed to achieve a non-Hermitian exceptional point (eigenvalue and eigenvector coalesce for $S$ at $r_b = 0$ in this case), i.e., UZR. For an example $(l_1, h) = (191 \text{ nm}, 16 \text{ nm})$, denoted by a star in Fig. 3(a), the $|t|$, $|r_f|$ and $|r_b|$ frequency spectra are plotted in Fig. 3(b) as symbols with UZR occurring at around 296 THz for this case. We note that the UZR is found here around one resonance mode and there is another resonance mode at around 400 THz without the UZR effect, which acts as a comparison in considering quantum interference later.

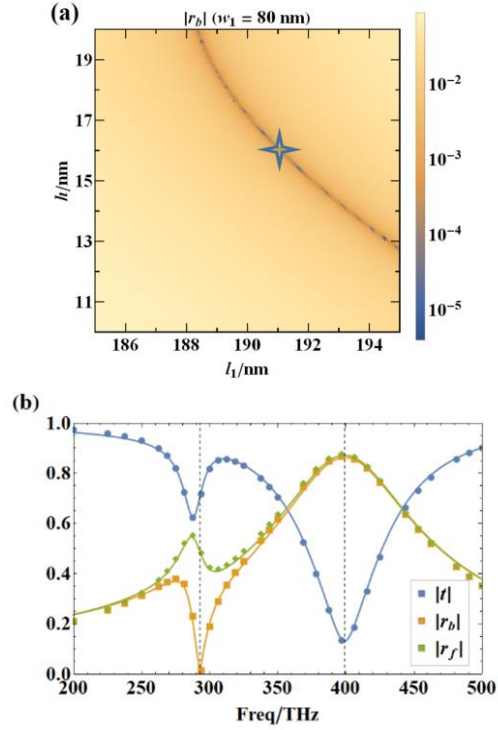

FIG. 3. (a) The minimum of $|r_b|$ around resonance frequency. The black curve indicates a family of UZR cases, and the star denotes the case shown in (b). (b) Spectrum of $S$ parameters for metasurface with $l_1$ = 191nm, $w_1$ = 80nm, and $h$ = 16nm. The numerical simulation (symbols) and the fitted CMT (solid lines) both show nearly zero reflectance of backward direction at around 296 THz, denoted as vertical dashed line. At another resonance, the transmission is greatly reduced while the reflection amplitudes on both sides rise (dashed line).

The classical scattering matrix of the family of UZR metasurfaces is generally non-unitary. Here, we adopt the ancilla-mode formulation to calculate the quantum state transformation, in which the classical 2x2 $S$ matrix is expanded into a 4x4 unitary matrix with two ancilla modes corresponding to the two loss channels of the metamaterial atoms [39,40]. In our case, $S = \{\{t, 0\}, \{r, t\}\}$ and if we have $|t|^2 = 1 - |r|$ satisfied, one ancilla mode can be dropped as there is a combination of input wave $\{|r|, t^*r\}$ for the two ports to excite no absorption. In this case, the 3x3 unitary Matrix with one ancilla mode (the 3$^{rd}$ index for the loss channel) becomes

$$S = \begin{pmatrix} t & 0 & -r^*t^2/\left(|t|^2\sqrt{|r|}\right) \\ r & t & t\sqrt{|r|} \\ t^*r/\sqrt{|r|} & -\sqrt{|r|} & |t|^2 \end{pmatrix}.$$

Now, suppose we consider quantum interference when we have a $|11\rangle$ input to the metasurfaces, we have

$$|110\rangle = \hat{a}_1^\dagger \hat{a}_2^\dagger |0\rangle$$

$$\rightarrow \left(t^*\hat{b}_1^\dagger + r^*\hat{b}_2^\dagger + t|r|^{-\frac{1}{2}}r^*\hat{b}_3^\dagger\right)\left(t^*\hat{b}_2^\dagger - |r|^{\frac{1}{2}}\hat{b}_3^\dagger\right)|0\rangle$$

$$= (|t|^2 - |r|)|r|^{-\frac{1}{2}}r^*|011\rangle - \sqrt{2}tr^*|002\rangle$$

$$-|r|^{\frac{1}{2}}t^*|101\rangle + t^{2*}|110\rangle + \sqrt{2}t^*r^*|020\rangle,$$

where the 3$^{rd}$ index in a "ket" is for the number of photons going into the absorption channel and $\hat{a}_i^\dagger$ ($\hat{b}_i^\dagger$) is the photon creation operation for mode $i$. The first term accounts for the quantum interference term into output $|01\rangle$ while the third term is into output $|10\rangle$ without quantum interference. We solve $|t|^2 - |r| = 0$, together with the single loss-channel condition $|t|^2 = 1 - |r|$. The condition of one-sided destructive quantum interference becomes

$$|t| = 1/\sqrt{2}, \qquad |r| = 1/2. \tag{1}$$

For general non-unitary scattering matrices (with only losses) other than the above values, one can follow the full ancilla modes approach [40] to obtain the probability for 1 photon at port 1 with 0 photon at port 2 or 0 photon at port 1 and 1 photon at port 2 are governed by

$$P_Q(0,1) = P_C(0,1) - t_f^* t_b^* r_f r_b$$
$$- t_f t_b r_f^* r_b^* - 2|t_b|^2 |r_f|^2, \tag{2}$$
$$P_Q(1,0) = P_C(1,0) - t_f^* t_b^* r_f r_b$$
$$- t_f t_b r_f^* r_b^* - 2|t_f|^2 |r_b|^2. \tag{3}$$

The difference to their classical counterparts $P_C(0,1) = |t_b|^2 A_f + |r_f|^2 A_b$ and $P_C(1,0) = |r_b|^2 A_f + |t_f|^2 A_b$ contributes to quantum interference where $A_f$ and $A_b$ are absorption coefficient in forward and backward directions respectively. When $r_b = 0$, the quantum interference in $P_Q(1,0)$ is turned off. Probabilities for other outputs are listed in Appendix I.

From the spectra shown in Fig, 3(b), we calculate quantum probabilities for different outputs from $|11\rangle$ input, as well as the classical probabilities. Fig. 4a shows the probabilities for $|10\rangle$ and $|01\rangle$ outputs (symbols for results calculated from full-wave simulation and lines from coupled mode model representation). At the UZR point (dashed line at 296 THz), $P_Q(1,0)$ is equal to $P_C(1,0)$ indicating the disappearance of quantum interference on the reflectionless side. On the other hand, $P_Q(0,1)$ and $P_C(0,1)$ differs a lot, $P_Q(0,1)/P_C(0,1) \approx 0.058$, due to the proposed one-sided destructive quantum interference. Despite the loss introduced from resonance, the destructive quantum interference effect eliminates the $|01\rangle$ output. In comparison, at another resonance around 400 THz, the detection probabilities for $|01\rangle$ and $|10\rangle$, no matter quantum or classical, are high because without the proposed condition (Eq. (1)) being satisfied. Fig. 4b shows the detection probabilities for $|20\rangle$ and $|02\rangle$ outputs as well. At the UZR point, the probability for output $|20\rangle$ reaches zero since the backward incident photon cannot be reflected. While the quantum probability for $|02\rangle$ still differs from its classical counterparts by a factor 2, resulting from a quantum effect, the mutual stimulation of two photons in the same state [41]. The asymmetric reflection of metasurface is inherited into the asymmetric interference effect in the quantum regime. The unidirectional reflection would eliminate the interference on the reflectionless side, while on the other side interference maintains tunable. Furthermore, with the help of the metasurface, we reach destructive interference of $|01\rangle$ output for $|11\rangle$ input while keeping the other side free of interference.

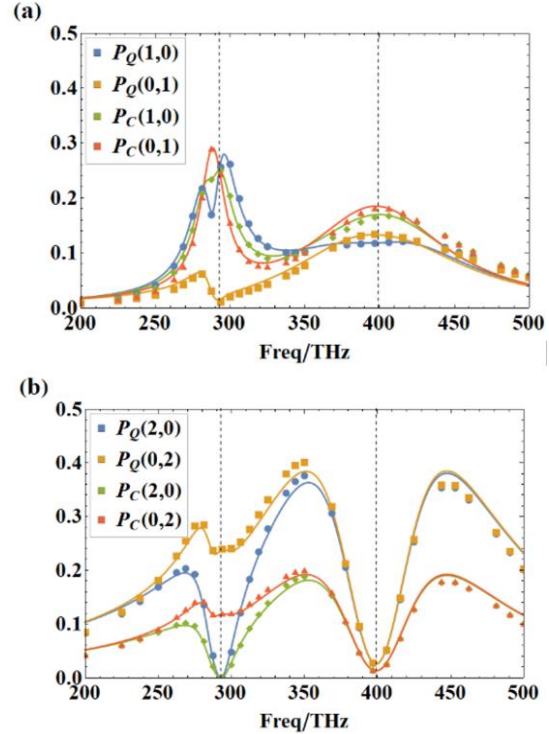

FIG. 4. Calculating the classical and quantum detection probability for output states (a) with only one photon and (b) with two photons from $|11\rangle$ input at different frequencies. Vertical dashed lines denote the resonance frequencies shown in Fig. 3b.

The proposed one-sided destructive quantum interference can also be implemented with conventional optical elements.

Using singular value decomposition [40], we can decompose the designed scattering matrix, $S = \frac{1}{2}\begin{pmatrix} \sqrt{2}e^{i\phi_1} & 0 \\ e^{i\phi_2} & \sqrt{2}e^{i\phi_1} \end{pmatrix}$, where $\phi_1$ and $\phi_2$ can be arbitrary, into two unitary matrices and a diagonal matrix denoting loss (see Appendix II). As shown in Fig. 5, the phase shifters are for adjusting the phase information of the scattering matrix. The beamsplitter (BS) 1 and 3 are two one-third BS corresponding to the two unitary matrices. The BS 2 is a 25/75 BS to introduce loss by interfering with ancilla vacuum state 3, while there is no loss in another channel thus no BS in this arm. While this decomposition into conventional optical elements can be used to realize the proposed one-sided destructive quantum interference, the tailor-made metasurface provides a more compact platform with subwavelength thickness and greatly simplifying the setup.

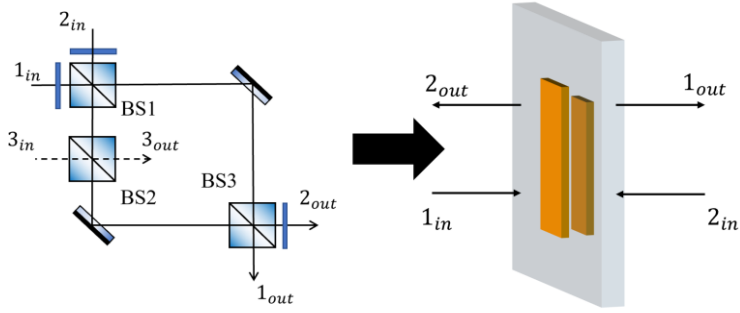

FIG. 5. The proposed one-sided destructive quantum interference can be implemented with the scheme using beamsplitters (BS) and now can be replaced by our designed metasurfaces to simplify the setup. Phase shifters are placed at the ports for phase adjustment. BS 1 and 3 are one-third BS while BS 2 is a 25/75 BS to introduce loss in channel 2 with vacuum mode 3.

In conclusion, we have proposed one-sided destructive quantum interference effect. This effect is from the non-Hermiticity of the system and can be realized by a bianisotropic metasurface with unidirectional zero reflection at exceptional point. Through this demonstration, tailor-made metasurfaces display their potential in discovering novel quantum interference effect and provide a more compact platform for quantum information processing.

Acknowledgment

The work is supported by Research Grants Council of Hong Kong through projects C6013-18G, 16304520 and AoE/P-502/20.